\documentclass[aps,twocolumn]{revtex4}

\usepackage{graphicx}
\usepackage{dcolumn}
\usepackage{bm}

\begin{document}

\title{Design and Simulation of THz Quantum Cascade Lasers}
%

\author{R\"udeger K\"ohler}
\address{NEST-INFM and Scuola Normale Superiore, Piazza dei Cavalieri 7, 56126 Pisa, Italy}
\author{Rita C. Iotti} 
\address{INFM and Dipartimento di Fisica, Politecnico di Torino, Corso Duca 
degli Abruzzi 24, 10129 Torino, Italy}
\author{Alessandro Tredicucci}
\address{NEST-INFM and Scuola Normale Superiore, Piazza dei Cavalieri 7, 56126 Pisa, Italy }
\author{Fausto Rossi}
\address{INFM and Dipartimento di Fisica, Politecnico di Torino, Corso Duca 
degli Abruzzi 24, 10129 Torino, Italy }
%
%

\begin{abstract}
Strategies and concepts for the design of THz emitters based on the quantum cascade scheme are analyzed 
and modeled in terms of a fully three-dimensional Monte Carlo approach; this allows for
a proper inclusion of both carrier-carrier and carrier-phonon scattering mechanisms. 
Starting from the simulation of previously published far-infrared emitters, where no 
population inversion is achieved, two innovative designs are proposed. The first one follows 
the well-established chirped-superlattice scheme whereas the second one employs a 
double-quantum well superlattice to allow energy relaxation through optical phonon 
emission. For both cases a significant population inversion is predicted at 
temperatures up to 80 K.
\end{abstract}
\maketitle
\begin{figure} 
\begin{center} 
\resizebox{0.85\columnwidth}{!}{
  \includegraphics*{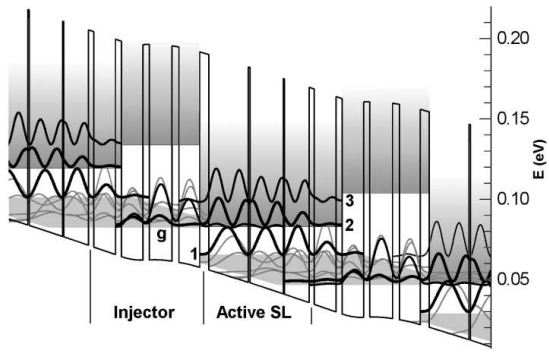}}
 \end{center}
\caption{
Conduction band energy diagram of the 'chirped' superlattice under an average electric field of 
3.5 kV/cm. The layer thickness (in nm) are, from left to right, starting from the injection 
barrier \textbf{4.3}/ 18.8/ \textbf{0.8}/ 15.8/ \textbf{0.6}/ 11.7/ \textbf{2.5}/ 10.3/ 
\textbf{2.9}/ 10.2/ \textbf{3.0}/ 10.8/ \textbf{3.3}/ 9.9, where Al$_{0.15}$Ga$_{0.85}$As layers are in
bold face and the 10.2 nm wide GaAs well is doped $4 \times 10^{16}\textrm{cm}^{-3}$. Also shown are the moduli 
squared of the wavefunctions; the optical transition occurs between states 2 and 1.
}
\label{fig:SL_Band} \end{figure}

\newpage
Since the first demonstration of the Quantum Cascade (QC) laser in 1994\cite{QCL}, its performance has
experienced tremendous improvement\cite{IEEE,Peltier}, and the range of emission wavelengths has been 
continuously extended\cite{3micron,17micron2,19micron,21micron}. At present, 
the longest wavelength QC lasers operate at $\lambda \sim 21.5\,\, \mu \textrm{m}$ and 
$24\,\, \mu \textrm{m}$\cite{21micron}, still above the LO-phonon energy threshold of
the host material ($\hbar\omega_{LO} \sim 34\,\, \textrm{meV}$ in InGaAs,
$\hbar\omega_{LO} \sim 36\,\, \textrm{meV}$ in GaAs). 
Although lasing at longer wavelength has not been observed yet, electroluminescence in the 
THz region of the 
spectrum has already been detected from a variety of QC structures\cite{87micron,MIT1,Gornik} as well as 
from other quantum well devices\cite{Helm,FIRemis}, stimulating a number of theoretical 
proposals\cite{Harrison}.
These strong efforts aimed at fabricating a THz semiconductor laser are mainly driven by 
the lack of compact, convenient solid-state sources operating at THz frequencies, despite the many 
possible applications in wireless communications, medical imaging, security screening, etc.\\
\indent
As prototypical design for THz emitters, we shall consider the GaAs/AlGaAs QC structure proposed by 
Rochat \emph{et al.}\cite{87micron}.
This can be regarded as a scaled down version of the conventional mid-infrared QC design, 
based on a vertical optical transition (see Fig. 1 in Ref. \onlinecite{87micron} for its band diagram). 
This implementation is characterized by a narrow injector miniband, chosen to avoid scattering of 
electrons from the upper state through LO-phonon emission and cross-absorption of the emitted light. 
Moreover, it features a relatively small electron tunneling between the states in the active 
region and in the injector.
As stressed before, while electroluminescent devices were reported by several 
groups\cite{87micron,FIRemis}, to date no population inversion could be achieved. In fact, 
several problems inherent to this design undermine the functionality of the structure.
First, the width of the miniband forbids the use of LO-phonon emission as the
main mechanism to deplete the lower state of the lasing transition.
Second, the small miniband dispersion and low tunneling probability also restrict efficient electrical 
transport to a narrow range of currents and voltages, limited by the onset of negative differential resistance. 
Third, the small energy separation between the levels in the active region and the quasi-Fermi level of the 
injector electrons makes the device very sensitive to hot-carrier effects ('backfilling' of 
the upstream active region).\\
\indent
In order to gain a better understanding of these limiting factors, we have applied a 
global simulation scheme, recently proposed in Refs. \onlinecite{Rita2,Rita3}, to the prototypical
structure of Rochat \emph{et al.}\cite{87micron}. 
Such a technique consists of a Monte Carlo sampling of a coupled set of fully three dimensional
(3D) Boltzmann-like equations. To properly model energy relaxation as well as carrier thermalization,
we included both carrier-phonon ({\it c-p}) and carrier-carrier ({\it c-c})
interaction mechanisms. Thanks to periodic boundary conditions, we are able to get
the device current-voltage characteristics as well as its gain spectrum without resorting to
phenomenological parameters.\\
\indent
At 80 K, under the design electric field of 2.7 kV/cm and assuming a sheet density of 1.2 $\times 10^{10}$
$\textrm{cm}^{-2}$, our simulated experiments give an electron density in the upper laser level 2 of about 
$8.5 \times 10^{8}\, \textrm{cm}^{-2}$, compared to the 
$3.5 \times 10^{9}\, \textrm{cm}^{-2}$ in the lower state 1. These values are far from 
the transparency condition of the intersubband transition, as already inferred 
from experimental data. This is not a consequence of rapid non-radiative depopulation of 
state 2 but rather of a slow extraction of electrons from level 1 into the downstream injector region. 
Indeed our simulation shows rather a long lifetime \cite{lifetime} of level 2, 
$\tau_{2}=2.2\, \textrm{ps}$, mainly determined by {\it c-p} scattering of electrons with high in-plane 
momentum to subband 1 ($\tau_{21} = 2.7\, \textrm{ps}$). 
On the other hand for level 1 
we compute a significantly longer lifetime $\tau_{1} = 14.5\, \textrm{ps}$ resulting from almost comparable 
{\it c-c} and {\it c-p} contributions. 
In reality, the electron dynamics is dominated by very fast {\it c-c} scattering with a 
quasi-resonant level (1' in Ref. 7). However, such a process acts in both directions 
($1\to 1'$ and $1'\to 1$) with nearly equal rates, thus giving only a marginal contribution
to the electron depletion of state 1. From the simulation we also obtain the value 
of the operational current density, 
$50 \textrm{A/cm}^{2}$, in very good agreement with the measured one at this bias field.\\
\indent
Starting from these considerations we have designed two structures tuned for emission at 
$\lambda \sim 69\,\, \mu\textrm{m}$ that can overcome the limitations associated with the above 
prototypical design. Both structures are based on a vertical-transition configuration, which is known to 
lead to larger dipole matrix elements and narrower linewidths. The first scheme employs a 
conventional chirped-superlattice design\cite{HighSL}, which allows flat minibands 
in the active region without requiring large 
concentrations of dopants. The band diagram under an average applied electric field 
of 3.5 kV/cm is shown in Fig.~ \ref{fig:SL_Band}. Its operating strategy is 
based on the same concepts successfully implemented for $\lambda \sim 17 - 24\,\, \mu\textrm{m}$ QC 
lasers\cite{17micron2,19micron,21micron}.
In order to minimize the density of electrons in the lower laser subband 1 we employ 
a dense miniband with seven subbands, which provides a large phase space where electrons 
scattered either from subband 2 or directly from the injector can spread. 
The miniband dispersion is chosen as large as possible compatibly with the need of avoiding 
cross-absorption. This suppresses thermal backfilling, and provides a large operating range of currents and 
voltages. Also in this structure, however, energy relaxation within the first miniband appears to be hindered 
by the lack of final states with appropriate energy to allow for LO-phonon emission. Nevertheless, 
carrier-carrier interactions may beneficially operate as an activation mechanism, which can provide 
sufficient in-plane momentum for the electrons to open a scattering path via optical phonon to the 
lower subbands. This can be clearly seen from Tab. 1, where we report the simulated 
steady-state distribution of carriers at 80 K in the various subbands, with and without {\it c-c} scattering 
in the simulation. 
As expected, in the absence of {\it c-c} interaction, the carrier density remains high in the upper 
states of the injector miniband and particularly 
in subband 1. 
On the other hand, in the presence of {\it c-c} scattering, electrons efficiently relax into the lower 
states of the injector (g,A), from where they are readily transferred into the upper laser 
state of the following period. In the latter case a population inversion of 
$\Delta n = 1.1 \times 10^{9}\textrm{cm}^{-2}$ results between subbands 2 and 1:
compared to the previous case, the number of electrons of subband 1 is decreased by 40\% 
whereas that of state 2 increases by a factor of four. Calculated lifetimes are  
$\tau_{2}=0.8\,\textrm{ps}$, $\tau_{21}=8.3\,\textrm{ps}$, $\tau_{1}=2.2\,\textrm{ps}$\cite{Note1} 
with $\tau_{2}$ and $\tau_{21}$ dominated by {\it c-p} scattering processes. 
Note that $\tau_{21}$ is much greater than $\tau_{2}$ 
due to the superlattice nature of the active region. This implies that $\tau_{1}<\tau_{21}$, which is the 
necessary condition for population inversion. Furthermore, the number of carriers escaping via subband 
3 is negligible.\\
\indent
In order to further improve the extraction rate from the lower laser state 1, we designed another 
structure which allows for a direct use of {\it c-p} scattering as the depleting mechanism 
without compromising the lifetime of the upper state 2. 
This is achieved by using a double-quantum well superlattice (DQW-SL)\cite{Mike}, 
which leads to a split miniband in the injector. 
By a careful design of the DQW-SL it is 
possible to separately adjust the sub-minibands dispersion and separation, where the latter can be chosen 
to match the LO phonon energy. Thereby we avoid cross-absorption and, at the same time, recover the possibility 
of using the efficient direct scattering by LO-phonons as the main relaxation process. 
To this end, we have also tried to delocalize the lower laser state 1 over both the active region and the 
following injector. Although this leads to a slight reduction of the dipole matrix element 
with the upper laser state, it increases  significantly the LO-phonon assisted transitions
to the injector subbands, in particular A, B, g. 
In that way the lifetime of state 1 is kept as short as 0.3 ps, 
resulting from both {\it c-c} and {\it c-p} scattering. In order to preserve the long 
lifetime of the upper laser level 2, the latter is well confined in only two wells and is 
 spatially separated from the injector. 
This keeps the scattering probability of electrons out of level 2 rather low, leading to a 
relatively long lifetime of 1.1 ps ($\tau_{21}= 1.9\, \textrm{ps}$), which is
dominated by {\it c-p} scattering. As reported in Tab. 2, population inversion in 
this structure is achieved even without taking into account {\it c-c} scattering. This 
confirms the good functionality of the phonon-split sub-minibands. In the presence of
{\it c-c} scattering, the population of the lower laser level increases due to faster scattering 
from level 2 to 1. However, this is compensated by a stronger injection from subband g 
into subband 2, resulting in an increased population of the latter.
The population inversion amounts to about 10 percent of the total sheet density per period.
Note that, owing to the larger voltage drop per period, thermal 'backfilling' of carriers is only 
a minor problem, thereby allowing for a higher doping in this second design.\\
\indent
Considering the dipole matrix elements of 7.8 nm and 4.5 nm computed for the optical transitions in the
two structures, respectively, and assuming a linewidth of 2 meV as experimentally detected at similar 
wavelengths\cite{87micron}, we estimate, on the basis of the above populations, gain coefficients of 
about 31 cm$^{-1}$ and 110 cm$^{-1}$. These values compare favorably to the optical losses 
(approximately 50 cm$^{-1}$) measured in double-surface plasmon waveguides at THz frequencies\cite{waveguide},
and indicate that laser action with such active region designs is a realistic goal.\\
\indent
In summary, we have proposed two QC structures designed for THz lasing.
Our theoretical analysis of their performance is based on a global simulation scheme
which allows for a proper inclusion of the most relevant scattering mechanisms.
The proposed designs both exhibit significant population 
inversion up to 80 K. This, together with the large transition dipoles, 
is expected to lead to high optical 
gain. Finally, it should be noted that our simulation predicts current densities of more than 1 kA/cm$^{2}$
which is roughly one order of magnitude higher than the current density measured in the 
prototypical design of Ref. \onlinecite{87micron} before the onset of negative differential
resistance. We attribute this effect to the improved carrier relaxation and higher 
tunneling efficiency of our structures.\\
\indent
This work was supported in part by the European Commission through the FET project WANTED.
R. K. acknowledges support by the C.N.R. The authors would like to thank Fabio Beltram for helpful discussions
and a careful reading of the manuscript.


\begin{table}
\caption{Population of the individual levels in the 'chirped' SL, calculated at 80 K with only carrier-phonon 
(c-p) scattering and with both c-p and carrier-carrier (c-c) scattering. Letters refer to the injector states 
comprising the first miniband in Fig.~ \ref{fig:SL_Band}. g represents the injector ground state while capital 
letters label the other levels in ascending order.}
\begin{ruledtabular}
\begin{tabular}{cccc}
Subband & Energy & Pop. (c-p only)              &     Pop. (c-p \& c-c)\\
 index  & (meV)  & (10$^{9} \textrm{cm}^{-2}$) & (10$^{9} \textrm{cm}^{-2}$)\\
\hline
3  & 51.7  & 0.02  & 0.06  \\ \hline
2  & 36.1  & 0.94  & 3.99  \\ \hline
1  & 17.9  & 4.72  & 2.83  \\ \hline
F  & 14.1  & 7.13  & 4.72  \\ \hline
E  & 11.0  & 8.08  & 6.08  \\ \hline
D  & 7.7   & 7.34  & 5.45  \\ \hline
C  & 4.5   & 5.14  & 5.98  \\ \hline
B  & 3.2   & 1.99  & 4.62  \\ \hline
g  & 0     & 6.19  & 7.87  \\ 
\end{tabular}
\end{ruledtabular}
\label{tab:table1}
\end{table}



\begin{table}
\caption{Population of the individual levels in the DQW SL, calculated at 80 K with only carrier-phonon (c-p)
scattering and with both c-p and carrier-carrier (c-c) scattering. Subbands are named with the same 
convention as in table \ref{tab:table1}.}
\begin{ruledtabular}
\begin{tabular}{cccc}
Subband & Energy & Pop. (c-p only)              &     Pop. (c-p \& c-c)\\
 index  & (meV)  & (10$^{10} \textrm{cm}^{-2}$) & (10$^{10} \textrm{cm}^{-2}$)\\
\hline
2  & 63.4  & 0.27   & 1.22  \\ \hline
1  & 45.5  & 0.07   & 0.46 \\ \hline
E  & 42.6  & 0.02   & 0.27  \\ \hline
D  & 37.1  & 0.05   & 0.27  \\ \hline
C  & 28.7  & 0.27   & 0.27  \\ \hline
B  & 9.2   & 1.22   & 1.16  \\ \hline
A  & 5.9   & 0.82   & 1.43  \\ \hline
g  & 0     & 5.24   & 2.99  \\ 
\end{tabular}
\end{ruledtabular}
\label{tab:table2}
\end{table}

\begin{figure} 
\begin{center} 
\resizebox{0.85\columnwidth}{!}{
  \includegraphics*{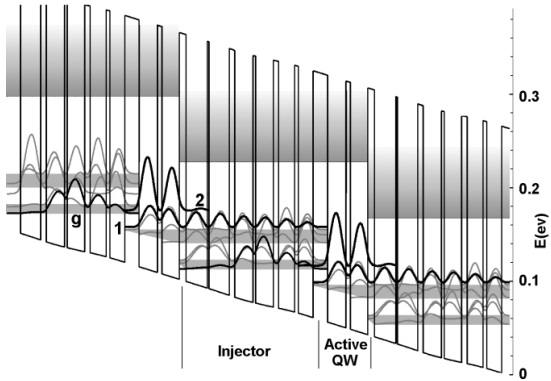}}
 \end{center}
\caption{
Conduction band energy diagram of the DQW-SL under an average electric field of 8.7 kV/cm. 
The layer thickness (in nm) are, from left to right, starting from the injection barrier
\textbf{3.5}/ 6.8/ \textbf{1.7}/ 6.4/ \textbf{2.5}/ 7.9/ \textbf{0.6}/ 7.5/ \textbf{2.0}/ \underline{6.8}/
\textbf{1.0}/ \underline{6.5}/ \textbf{2.0}/ 5.9, \textbf{1.4}/ 5.5, where Al$_{0.3}$Ga$_{0.7}$As layers are in bold face and doped 
(6 $\times$ 10$^{16}$ cm$^{-3}$) GaAs layers are underlined. The Al concentration in this design, and thus the 
barrier height, is raised compared to the previous design in order to better accomodate the split miniband.
}
\label{fig:Tredi1} \end{figure}

\end{document}